\documentclass[aps, pra, superscriptaddress,twocolumn]{revtex4-2}

\usepackage{xcolor}
\usepackage{amssymb}
\usepackage{bbm}
\usepackage{mathrsfs}
\usepackage{verbatim}
\usepackage{dsfont}
\usepackage{graphicx}
\usepackage{bm}
\usepackage{amsmath}
\usepackage{amssymb}
\usepackage{mathtools}
\usepackage[colorlinks=true, allcolors=black]{hyperref}
\usepackage{tikz}
\usetikzlibrary{decorations.pathreplacing}
\usepackage{float}

\DeclarePairedDelimiter\ceil{\lceil}{\rceil}

\newcommand\im{i}
\newcommand{\identity}{{\mathbb I}}

\newcommand\erg{\mathcal{E}}
\newcommand\ro{\hat{\rho}}

\newcommand\tr{\mathrm{Tr}}

\newcommand\W{\mathcal{W}}

\newcommand\e{\mathfrak{e}}
\newcommand\g{\gamma}

\newcommand\hc[1]{#1^\dagger}
\newcommand\proj[1]{|#1\rangle\!\langle#1|}

\newcommand\ketbra[2]{|#1\rangle\!\langle#2|}

\newcommand\ketb[1]{{|#1\rangle}_B}
\newcommand\brab[1]{\prescript{}{B}{}\langle#1|}
\newcommand\ket[1]{|#1\rangle}
\newcommand\bra[1]{\langle#1|}

\newcommand\pass[1]{\hat{\pi}_{#1}}
\newcommand\gibbs{\hat{\tau}}

\newcommand\G{\mathcal{G}}
\newcommand\J{\mathcal{J}}

\newcommand\dephased[1]{\hat{\delta}_{#1}}
\newcommand\shuffled[1]{\hat{\sigma}_{#1}}

\begin{document}



\title{Generalized multilevel amplitude damping channels and their thermodynamic performances}

\date{\today} 
\author{Vito Vetrano}\email{vito.vetrano@sns.it}
\affiliation{Scuola Normale Superiore, 56126 Pisa, Italy} 
\author{Vittorio Giovannetti}\email{vittorio.giovannetti@sns.it}
\affiliation{Scuola Normale Superiore, 56126 Pisa, Italy} 
\author{Vasco Cavina}\email{vasco.cavina@sns.it}
\affiliation{Scuola Normale Superiore, 56126 Pisa, Italy} 


\begin{abstract}
We introduce a new class of quantum channels, the Generalized Multilevel Amplitude Damping (GMAD) channels, to model noise and decoherence effects in a qudit coupled to a thermal environment.
The degradation of energetic resources under GMADs is investigated by evaluating work functionals and ergotropic capacitances, with particular attention to the coherent and incoherent contributions to ergotropy, for which we introduce new quantifiers.
Our analysis sheds light on how to optimally prepare a qudit in a thermal environment in order to preserve its value from the perspective of work extraction, and reveals several counterintuitive phenomena: the ergotropic capacitance of a GMAD channel is not monotonic in the temperature of the environment; moreover, iterating the map can lead to crossings between ergotropic functionals at different temperatures, indicating the presence of a Markovian Mpemba effect.

\end{abstract}


\maketitle

\section{Introduction}

In recent years, the quest for achieving dense and controllable quantum architectures has led to a renewed interest in refined descriptions of noise models.
Despite the impressive amount of general theory \cite{gyongyosi2018survey}, an explicit characterization of exact noise models, i.e., an in-depth study of the capacities of specific quantum channels, remains relatively unexplored. This statement is even more true when we leave the more consolidated ground of information capacities of quantum channels and look toward energetic capacities, which measure the capability of a channel to allow a safe encoding of energy in quantum states in order to make this energy usable and retrievable later \cite{francica2020quantum,energylines,QWC}. The importance of the preservation and transfer of energy in quantum architectures has been recognized recently with the advent of quantum thermodynamics \cite{alicki2019introduction, deffner2019quantum, vinjanampathy2016quantum} and it is expected to play an important role in the path toward efficient quantum computation \cite{auffeves2022quantum, castellano2025exact, gois2024towards, silva2023classical} and toward efficient quantum energy storage \cite{campaioli2018quantum, quach2022superabsorption, canzio2025single}.

From the formal point of view of quantum information theory, the most common approach is to model noise through a quantum channel, i.e., a completely positive and trace-preserving (CPTP) map acting on the system of interest \cite{breuer2002theory}.
In this context, single-qubit channels are the most extensively studied \cite{nielsen2010quantum, khatri2020information,giovannetti2005information}, while a detailed characterization of multi-level models, such as the Multilevel Amplitude Damping (MAD) channel and its resonant version, has been achieved only recently \cite{chessa2021quantum, cocciaretto2026quantum, chessa2023resonant}.
In this manuscript, we focus on an extension of the MAD model called the Generalized Multilevel Amplitude Damping (GMAD) channel, with the main purpose of extending the existing theory to situations in which the thermal excitations coming from the environment are explicitly taken into account.
In addition, since we consider an arbitrary number of levels, we can account for dissipation and decoherence effects that go beyond the single-qubit scenario (and are suited to more general situations like globally-acting noise \cite{menta2025globally} and noise acting on qudits).
To evaluate the performance of GMADs from a thermodynamic perspective, we quantify the energy that can be extracted from their output states, under suitably chosen constraints on the input. The extractable energy is measured in terms of the ergotropy, i.e., the maximum energy that can be retrieved via unitary operations \cite{hatsopoulos_unified_1976, allahverdyan2004maximal}.

The channels studied display several counterintuitive phenomena. Despite modeling physical systems in contact with thermal environments—which typically leads to the degradation of thermodynamic resources such as free energy \cite{brandao2015second, ng2019resource}—GMADs can increase ergotropy, suggesting that dissipative dynamics may exhibit regimes that can be exploited for work extraction. Moreover, when varying the temperature of the environment, GMADs at higher temperatures can sometimes perform worse than their lower-temperature counterparts in preserving ergotropic resources. This behavior hints at physics analogous to the Markovian Mpemba effect \cite{lu2017nonequilibrium, carollo2021exponentially}.
We attribute this effect to the interplay of both coherent and incoherent mechanisms.

The paper is organized as follows. In Sec.~\ref{sec:defpro} we introduce the GMAD; in Sec.~\ref{sec:ergfunc} we review the definition of ergotropy and channel ergotropy and introduce the quantities used throughout the paper. In Secs.~\ref{sec:enact} and \ref{sec:degco} we assess the thermodynamic performance of GMADs and discuss the counterintuitive effects described above. Finally, in Sec.~\ref{sec:concl} we present our conclusions.

\section{Generalized Multilevel Amplitude Damping Channels}
\label{sec:defpro}
We consider a $d$-level system with Hamiltonian $\hat{H} = \sum_{j=0}^{d-1} \epsilon_j \proj{j}$, that we will assume to be non-degenerate, in contact with an environment of inverse temperature $\beta$. 
To model its open dynamics, we introduce a new class of linear, completely positive, trace preserving (LCPT) maps, the Generalized Multilevel Amplitude Damping channels (GMADs), that we write in Kraus representation as
\begin{equation} \label{eq:GMADdef}
    \Phi(\ro) = \sum_{m=1}^{d-1} \sum_{\overline{i,j} = 0}^{m} 
    \hat{K}_{ij}^{(m)} \ro 
     \hat{K}_{ij}^{(m)\, \dag} + \Phi_{D}(\ro)
\end{equation}
where we used $\overline{i,j}$ as a contracted notation for $i \neq j$ and 
$\Phi_{D}$ is a CP map containing all the diagonal Kraus operators, as specified later in this section.
The off-diagonal Kraus operators are associated with simple jumps between the energy levels
\begin{equation} \label{eq:offkraus}
 \hat{K}_{ji}^{(m)} = \gamma^{(m)}_{ij}  \ket{i}\bra{j}, 
\end{equation}
and the coefficients $\gamma_{ij}^{(m)} \coloneqq u_{ij}^{(m)} r_{mj}$ are the elements of an $(m +1) \times (m+1) $ matrix such that
\begin{equation} \label{eq:unitdb}
    \sum_{k=0}^{m} u_{ik}^{(m)} u_{i'k}^{(m)\,*}= \delta_{ii'}\ , \qquad r_{mj} = Z^{-\frac{1}{2}} e^{-\frac{\beta}{2} \omega_{mj}}\ ,
\end{equation}
where $\omega_{ij} = \epsilon_i - \epsilon_j$ and $Z$ is a normalization constant fixed by the trace preservation condition of the channel, that is, such that $Z = \sum_{m >j} e^{-\beta \omega_{mj}} +1$.
The definition above is the simplest that ensures two fundamental properties. First, it admits a physical representation in which the system is unitarily coupled to a thermal environment via a coupling Hamiltonian that preserves the total energy of the system–environment composite. Second, in the limit $\beta \rightarrow \infty$, the off-diagonal Kraus operators reduce to the ones of a $d$-level MAD, and the off-diagonal Kraus structure of all MADs can be obtained in this way (see App. \ref{app:microGMAD}).
Along the same lines, the diagonal part of the channel is defined as 
\begin{equation} \label{eq:diagmad}
      \Phi_D(\ro) = \sum_{i=1}^{d-1} \sum_{j=0}^{i-1} \hat{K}_{ij}^{(D)}
     \ro 
     \hat{K}_{ij}^{(D)\, \dag} + \hat{K}_{00}^{(D)} \ro \hat{K}_{00}^{(D)\, \dag}
\end{equation}
where we introduced the following set of diagonal Kraus operators
\begin{align}
\notag
& \hat{K}_{ij}^{(D)} = r_{ij}\Big[\identity+(u_{jj}^{(i)}-1)\ketbra{j}{j}\Big],
\\  
& \hat{K}_{00}^{(D)}  = Z^{-\frac{1}{2}} \sum_{k=0}^{d-1} u_{kk}^{(k)}\ketbra{k}{k}.
\label{eq:diagkraus}
\end{align}
Notice that in the limit $\beta \rightarrow \infty$ we have $r_{ij} \rightarrow 0$, thus the only diagonal Kraus operator that survives is $\hat{K}_{00}^{(D)}$, the unique one appearing in the MAD \cite{chessa2021quantum}.
Although the number of Kraus operators presented here exceeds the optimal value, in Appendix \ref{app:minimalset} a minimal Kraus set for the GMAD is constructed.

The channel introduced above has many remarkable properties. 
It directly stems from its microscopic interpretation that the GMAD is a {\it thermal operation} \cite{brandao2015second}, and as such is also Gibbs-preserving
\begin{equation}
    \Phi(\gibbs) = \gibbs\ ,
\end{equation}
where $\gibbs$ is the Gibbs state associated with the system Hamiltonian.
Given its Kraus structure the GMAD is easily shown to decouple the dynamics of coherence from that of populations and to be {\it strictly incoherent} \cite{baumgratz2014quantifying, marvian2016quantum, winter2016operational}.
\begin{figure}
    \centering
    \includegraphics[width=\linewidth]{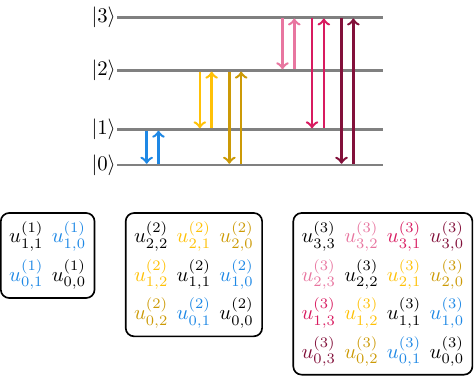}
    \caption{Qualitative sketch of the effect of the GMAD channel on a generic 4-level system.
    Following the definitions in Eq. \eqref{eq:GMADdef}, we can isolate three blocks of Kraus operators (corresponding to $m=1,2,3$) acting on the first $2, 3, 4$ levels, respectively.
    The properties of the coefficients $u_{ik}^{(m)}$
 appearing in the rates in Eq. \eqref{eq:offkraus} are made explicit in the bottom of the figure: for a fixed $m$ they can be arranged as entries of an $m+1$ dimensional unitary matrix. Jumps connecting the energy levels are depicted using distinct colors and hues, and the entries of the unitary matrices are highlighted using the same colors as the corresponding transitions.}
    \label{fig:sketch}
\end{figure}
The physically motivated definition allows us to encode relevant properties, such as {\it detailed balance conditions} 
 \cite{alicki1976detailed, fagnola2008detailed, soret2022thermodynamic},
 by adding some constraints to the parameters. 
More precisely, GMADs that originate from a microreversible dynamics $u_{ij}^{(m)} = u_{ji}^{(m)} $
have transition probabilities $P_{i|j} \coloneqq \sum_m |\gamma_{ji}^{(m)}|^2 $  that satisfy
\begin{equation} \label{eq:db}
\frac{P_{i|j}}{P_{j|i}} = \frac{\sum_m |\gamma_{ji}^{(m)}|^2}{\sum_{m} |\gamma_{ij}^{(m)}|^2 } = e^{- \beta \omega_{ij}}.
\end{equation}
We conclude by noting that the non-degeneracy of the Hamiltonian of the system is a fundamental requirement to obtain the simple structure of the off-diagonal Kraus operators in Eq. \eqref{eq:offkraus}.
If we drop this assumption, individual Kraus operators can induce multiple energy transitions simultaneously, and such a construction would correspond to a generalization of a zero-temperature model different from the MAD, namely the one considered in \cite{chessa2023resonant}, which lies outside the scope of the present work. 
We finally stress that, for the specific case of $d=2$, i. e., a qubit, the GMAD encompasses the generalized amplitude damping channel (GADC) \cite{khatri2020information, nielsen2010quantum} if we choose $u^{(1)}_{ij} \in \mathds{R}$ for all $i,j \in \{0,1\}$. 
\section{Ergotropic functionals}
\label{sec:ergfunc}
A functional of primary importance in the context of work extraction is {\textit{ergotropy}} \cite{hatsopoulos_unified_1976,allahverdyan2004maximal}. The ergotropy $\erg(\ro; \hat{H})$ of a state $\ro$ given the Hamiltonian $\hat{H}$ is the maximum work unitarily extractable from $\ro$:
\begin{equation}
    \label{eq:def-erg}
    \erg(\ro; \hat{H}) \coloneqq \max_{\hat{U}\in \mathbb{U}(d)}\left\{\tr[\ro \hat{H}]-\tr[\hat{U}\ro \hc{\hat{U}} \hat{H}]\right\}
\end{equation}
with $\mathbb{U}(d)$ the set of unitary transformations acting on a $d$-dimensional system; we will suppress the explicit $\hat{H}$ dependence where possible, to ease the notation. The maximum in \eqref{eq:def-erg} is attained for the unitary that sends $\ro$ into \textit{its passive state} $\pass{\rho}$, i.e. the diagonal state obtained by populating the energy eigenspaces $\{\ket{j}\}_j$, from bottom to top, using the state's eigenvalues $\{\lambda_j\}_j$, from the greatest to the lowest:
\begin{equation}
    \label{eq:pass-of-rho}
    \pass{\rho} = \sum_{j=1}^{d}\lambda_j^\downarrow \proj{j}\ ;\quad\erg(\ro)= \tr[(\ro -\pass{\rho})\hat{H}]\ .
\end{equation}
If an asymptotically large number of identical copies of the state is available, the \textit{total ergotropy} \cite{pusz1978passive, lenard1978thermodynamical} is suitable for quantifying the best performance of a work extraction task \cite{QWC}:
\begin{equation}
    \label{eq:ergtot}
    \erg_{tot}(\ro;\hat{H}) \coloneqq \lim_{n\to+\infty}\frac{1}{n}\erg(\ro^{\otimes n};\hat{H}^{(n)}) = \tr[(\ro-\gibbs_{\rho})\hat{H}]\ ,
\end{equation}
where $\gibbs_{\rho}$ is a thermal state of $\hat{H}$ with an inverse temperature $\beta^*_{\rho}$ such that $S(\gibbs_{\rho})=S(\ro)$, with $S$ the von Neumann entropy.

Since the coherence of a state (and the experimenter's ability to use it) can dramatically alter the amount of work that can be retrieved from it, it is convenient to introduce some analogues of the ergotropy to take this into account.
The \textit{incoherent ergotropy} was introduced in \cite{francica2020quantum} to quantify the maximum work extractable from a state using only the \textit{incoherent operations} $\mathbb{U}_{i}(d)$, i.e. the unitary transformations that do not change the coherence in the energy eigenbasis. Such operations correspond to a reshuffling of the energy eigenbasis, up to phase factors \cite{baumgratz2014quantifying}:
\begin{equation}
    \label{eq:def-Unc}
    \hat{V}\in \mathbb{U}_{i}(d) \iff \hat{V} = \sum_{l=0}^{d-1}e^{-i\varphi_l} \ketbra{R(l)}{l}\ ,
\end{equation}
where $R$ is some permutation acting on the indices $l$.
Hence, given a state $\ro$ and the Hamiltonian $\hat{H}$, the incoherent ergotropy $\erg_{i}(\ro, \hat{H})$ is:
\begin{equation}
    \label{eq:def-erg_nc}
    \erg_{i}(\ro; \hat{H}) \coloneqq \max_{\hat{V}\in \mathbb{U}_{i}(d)}\left\{\tr[\ro \hat{H}]-\tr[\hat{V}\ro \hc{\hat{V}} \hat{H}]\right\}\ .
\end{equation}
It is now not difficult to see (for a detailed proof see \cite{francica2020quantum}) that the incoherent ergotropy is equal to the ergotropy of the completely decohered state 
\begin{equation}
    \erg_{i}(\ro)= \erg(\dephased{\rho})\ ,
    \label{eq:erg_nc-is-erg-dephased}
\end{equation}
with $\dephased{\rho} = \sum_{j} p_j \proj{j}$, where $p_j$ are the diagonal elements of the density matrix in the energy eigenbasis.

From the relation $\mathbb{U}_{i}(d)\subset\mathbb{U}(d)$ we immediately get $\erg(\ro)-\erg_{i}(\ro)\geq 0$. Thus, it is possible to define the \textit{coherent ergotropy} as:
\begin{equation}
    \label{eq:def-erg_c}
    \erg_c(\ro; \hat{H})\coloneqq\erg(\ro)-\erg_{i}(\ro)=\tr[(\pass{\delta_{\rho}}-\pass{\rho})\hat{H}]\geq 0\ .
\end{equation}
To better characterize what $\erg_c$ corresponds to, let $\shuffled{\rho}:=\tilde V\ro\hc{\tilde V}$ indicate the state after the optimal permutation in equation \eqref{eq:def-erg_nc} is applied, i.e. the one with populations in the energy eigenstates reordered from greatest to lowest. Then, the coherent ergotropy can be expressed as \cite{francica2020quantum}:
\begin{equation}
    \erg_{c}(\ro)= \erg(\shuffled{\rho})\ .
\end{equation}
Hence, the coherent ergotropy of a state is equal to the ``leftover" ergotropy of the state after populations have been reordered.
Note that both the ergotropy and the incoherent ergotropy are convex functions with respect to their input state (see, for instance, the appendix of \cite{canzio2025extracting}).
Since $\mathcal{E}_c$ is the difference between two convex functions of $\ro$, it has no convexity properties in general; in particular, in Appendix \ref{app:Workcoh} it is shown that it is not even \textit{quasiconvex} \cite{greenberg1971review}.
Finally, it should be noted that, although coherence and the coherent ergotropy are tightly related, the latter is not a coherence monotone; this means that there are incoherent transformations \cite{baumgratz2014quantifying} that do not increase the coherence of the state, which however strictly increase $\erg_c$ (see the Appendix A of \cite{francica2020quantum} for a detailed example).

\subsection{Maximal Work Functionals}
All the functions introduced above characterize the maximum work yield of a system in a given state $\ro$, in different settings. 
Our goal is to assess the behavior of these functions under the action of the GMAD.
The problem is similar to the usual information theoretical setting of quantum communication theory, in which one seeks to study how much the information about a given state is retrievable after the action of a channel. We can thus borrow, following \cite{QWC, energylines}, the approach of quantum communication theory and apply it to our problem; this will lead to the definition of the so–called {\it work capacitances}.

Suppose that one has access to a collection of $n$ identical and independent systems.
The first agent, ``Alice”, encodes the energy $E$ in their $n$-partite state $\ro^{(n)}$, while a second agent, namely ``Bob”, receives the (possibly degraded) state and tries to extract work from it.
The noisy channel is a generic map $\Lambda$, such that the final state after the interaction is $\ro^{(n)}_{out}= \Lambda^{\otimes n}\big(\ro^{(n)}\big)$.
Fixing the input energy $E \geq 0$ and the number of copies $n$, one can define the following $n$-cells Maximal Ergotropic Functional (MEF) (see \cite{QWC} or Appendix \ref{app:workfunc}):
\begin{equation}
    \label{eq:ergchan}
       \mathcal{E}^{(n)}(\Lambda;E) \coloneqq \max_{\ro^{(n)}\,\in\,\mathfrak{I}^{(n)}_E}\ \erg(\Lambda^{\otimes n}(\ro^{(n)});\hat{H}^{(n)})\ ,\\
\end{equation}
where $\mathfrak{I}^{(n)}_E = \mathfrak{S}^{(n)}_E, \mathfrak{B}^{(n)}_E$ are the subsets of the system Hilbert space satisfying respectively the condition $ \tr[\ro^{(n)} \hat{H}^{(n)}] = E$ (in which case, we will talk of \textit{shell}-MEF), or $\tr[\ro^{(n)} \hat{H}^{(n)}] \leq  E$ (in which case, we will instead talk of \textit{ball}-MEF). Reasoning in an analogous way, to characterize the degradation of incoherent/coherent resources during the action of a quantum channel, we introduce the $n$-cells \textit{Maximal Incoherent} and \textit{Maximal Coherent Ergotropic Functionals}:
\begin{equation} \label{eq:ergchanco}
    \begin{gathered}
        \erg_{i,\,c}^{(n)}(\Lambda;E) \coloneqq \max_{\ro^{(n)}\,\in\,\mathfrak{I}^{(n)}_E}\ {\erg}_{i,\,c}(\Lambda^{\otimes n}(\ro^{(n)});\hat{H}^{(n)})\ .
    \end{gathered}
\end{equation}
As outlined in \cite{energylines}, the maximization over the sets $\mathfrak{I}^{(n)}_E$ in Eq. \eqref{eq:ergchan} can be restricted to the pure states only. It is straightforward to see that the same property holds for the functional $\erg_{i}^{(n)}(\Lambda;E)$ defined in Eq. \eqref{eq:ergchanco} (see App. \ref{app:Workcoh}), while it is not true for $\erg_{c}^{(n)}(\Lambda;E)$ in general.
In many cases, however, the optimal state could be challenging for Alice to prepare or for Bob to manipulate; therefore, it is reasonable to introduce versions of the MEFs in which the abilities of the sender or the receiver are reduced, as the \textit{Separable-Input Maximal Work Functionals} \cite{QWC}
where the maximization is performed over the separable input states in $\mathfrak{I}^{(n)}_E$, or MEFs in which the ergotropy is replaced by the {\it Local Ergotropy}, that is, the set of unitaries accessible to the receiver is limited to local operations on the single copies \cite{perarnau2015extractable}.
These quantities will appear occasionally in the following discussion. To avoid overloading this section and since their operational meaning is hopefully already clear, we collect their formal definitions in Appendix \ref{app:workfunc} for the reader’s convenience.
\subsection{Work Capacitances}
Work capacitances are used to describe the asymptotic capability of a channel to preserve energetic resources, measured in terms of the ergotropy of the final state.
To define capacitance, we prepare $n$ copies of the system and fix the amount of energy per copy $\e := E/n$.
Customarily, the Hamiltonian is renormalized such that $\e$ takes values in $[0, 1]$ and consequently $E \in [0, n]$. In this framework the \textit{Work Capacitance} reads \cite{QWC}:
\begin{equation}
    \label{eq:def-capacity}
    C_{\W}(\Lambda;\e) \coloneqq \limsup_{n \to \infty}\frac{\W^{(n)}(\Lambda;E=n\e)}{n}\ ;    
\end{equation}
where $\W$ is a placeholder for any of the work functionals maximized over $\mathfrak{B}_E^{(n)}$ introduced in the previous subsection.
The evaluation of these quantities for a generic open systems remains a hard problem, since super-additivity effects may take place \cite{QWC}.
Interestingly, there are some completely general results that connect the capacitances computed in the local and separable frameworks with the single-shot MWFs (i.e. the quantities defined in Eqs. \eqref{eq:ergchan}, \eqref{eq:ergchanco} for $n=1$).
First of all, it is easy to prove
\begin{align} \notag
 C_{\mathcal{W},loc,sep}(\Lambda;\e) & \leq  C_{\mathcal{W},sep}(\Lambda;\e) \ , 
 \\ C_{\mathcal{W},\,loc}(\Lambda;\e) & \leq C_{\mathcal{W}}(\Lambda;\e)\ ,
  \label{eq:WC-hierarchy}
\end{align}
where the suffix $loc, \, sep$ denote that the encoding is done by Alice on separable states and the work extraction is done by Bob using local operations, respectively.
The relation between $ C_{\mathcal{W},sep}(\Lambda;\e)$ and $C_{\mathcal{W},loc}(\Lambda;\e)$ is not immediately clear, but a further refinement of the hierarchy \eqref{eq:WC-hierarchy}, derived in \cite{WEroleofnonlocalresource} in the case of $\mathcal{W} = \mathcal{E}$ and extended in App. \ref{app:Workcoh} for $\mathcal{W} = \mathcal{E}_{i}$, shows that
\begin{equation}
   C_{\mathcal{W},loc,sep}(\Lambda;\e) = C_{\mathcal{W},loc}(\Lambda;\e) \leq C_{\mathcal{W},sep}(\Lambda;\e)\ ,
    \label{eq:WC-complete-hierarchy}
\end{equation}
and the first quantity is connected to single-shot ergotropy through the relation
\begin{equation}
    \begin{gathered}
      C_{loc,sep}(\Lambda;\e) =  \chi(\Lambda;\e) \coloneqq \sup_{\{p_j,\e_j\}} \sum_j p_j \erg^{(1)}(\Lambda;\e_j)\ ,
        \label{eq:def_chi}
    \end{gathered}
\end{equation}
with $\{p_j\}_j$ probabilities and $\e_j \in [0,1]\ \forall j$ (the interval is fixed by the normalization of $\e$) such that $ \sum_j p_j \e_j \leq \e\ $.
The result in Eq. \eqref{eq:WC-complete-hierarchy} establishes that, in the case of local work extraction by Bob, using separable or entangled input states gives the same asymptotic performances.
The result in Eq. \eqref{eq:def_chi} will be of particular interest for us: as the optimization in the r.h.s. can be worked out explicitly for some particular GMADs, this will allow us to easily compute the capacitances associated to such channels.

\section{Degradation of coherent and incoherent work in a three level GMAD}
\label{sec:degco}
We will now present a detailed analysis of the performance of the GMAD through the functionals defined in Sec.~\ref{sec:ergfunc}, focusing our attention on the qutrit ($d=3$) case. The Kraus operators of a GMAD acting on a qutrit have the structure introduced in Eqs. \eqref{eq:offkraus} and \eqref{eq:diagkraus}, 
and are fully characterized by the choice of two unitary matrices $u^{(2)}, u^{(1)}$.
To determine these matrices, we start from a microscopic, physically motivated model in which the qutrit is in contact with a thermal environment through an excitation preserving Hamiltonian of the form:
\begin{equation}
    \hat{H}_{I} = \sum_{k} g_k(\hat{\sigma}_{k,S}^{-}\ketbra{k}{0}_B + \hat{\sigma}_{k,S}^{+}\ketbra{0}{k}_B )\ ,
\end{equation}
where $k$ is a collective index running over all the possible ordered pairs $\{r s\},\ r > s$, of levels of the system, and $\hat{\sigma}^-_{rs,S} :=\ketbra{s}{r}_S $, $\hat{\sigma}^+_{rs,S} :=(\hat{\sigma}^-_{rs,S})^\dagger$ (see Appendix \ref{app:gmadqutrit} for a complete and explicit derivation of the GMAD Kraus operators from this interaction Hamiltonian). Despite its simple form, this Hamiltonian appears in several distinct physical contexts, modeling dipole-dipole interactions as well as beam splitter interactions and coherent cavity hopping \cite{tavis1968exact, loudon1974quantum, sirkina2025forstertransferquantumdots, hartmann2006strongly}.

\subsection{Shell-Maximal Ergotropic Functionals}
\begin{figure}
    \centering
    \includegraphics[width=1\linewidth]{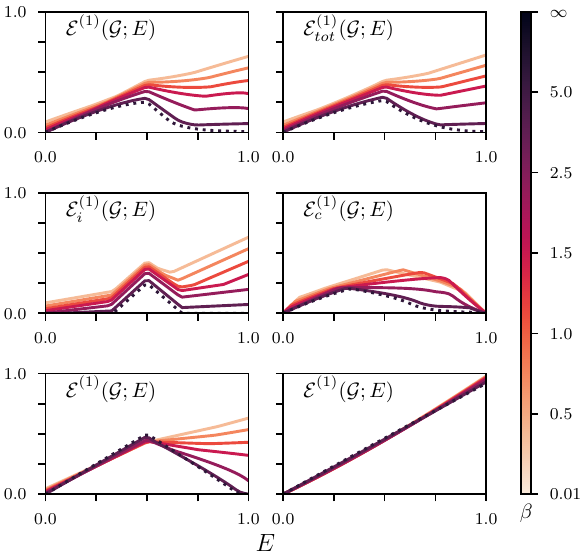}
    \caption{(From top left to center right) 1-cell shell-maximal work functionals for the channel $\G$ with $s_1=0.5$, $\bar{s}=0.99$, $\alpha_0=0.99$, as a function of the energy $E$, for different values of $\beta$. The dashed line represents the corresponding MAD channel ($\beta=+\infty$).
    (Bottom left) 1-cell shell-maximal ergotropy as a function of $E$ for the channel $\G$ with $s_1=0.01$, $\bar{s} = 0.99$, $\alpha_0=0.745$, and different values of $\beta$. The non-monotonicity of the maximal ergotropy in the temperature is evident.
    (Bottom right) 1-cell shell-maximal ergotropy as a function of $E$ for the channel $\G$ with $s_1=0.99$, $\bar{s} = 0.255$, $\alpha_0=0.5$, for different values of $\beta$. In this case, the maximal ergotropy almost coincides with the input energy.}
    \label{fig:erg-beta}
\end{figure}

Curves for the 1-cell shell-maximal ergotropy and shell-maximal total, incoherent and coherent ergotropies as a function of the input energy $E$ are plotted in Fig.~\ref{fig:erg-beta}, for a certain choice of the map parameters $(s_1, \bar{s},\alpha_0)$, introduced and motivated in Appendix \ref{app:gmadqutrit}, and defined as:
\begin{equation}
    \begin{gathered}
        s_1 \coloneqq \sin{(g_{10}t)}\,,\ \bar{s} \coloneqq \sin{(\bar{g}t)}\,,\ \alpha_0 \coloneqq g_{20}/\bar{g}\,,
        \\\bar{g}\coloneqq\sqrt{g_{20}^2+ g_{21}^2}\,.
    \end{gathered}
\end{equation}
In this case, all the functionals acquire their maximum on pure input states, including coherent ergotropy; this last fact is proved in Appendix \ref{app:Workcoh}.

The figure allows us to highlight several properties:
\begin{itemize}
    \item One of the first things to be noted is the non-monotonicity of the shell-maximal ergotropic functionals as a function of the energy (see top four panels in Fig.~\ref{fig:erg-beta}).
    The same non-monotonicity is also evident in the plots for shell-maximal incoherent and coherent ergotropies. In particular, the coherent ergotropy (center right panel in Fig.~\ref{fig:erg-beta}) is bound to be zero when $E$ equals one of the the extremities of the Hamiltonian spectrum, as the iso-energetic surface collapses to an eigenstate of $\hat{H}$, and the map $\G$ is strictly incoherent.
    
    \item Although all the functionals seem to increase with the temperature, this should not lead to the conclusion that this monotonicity holds in general. As showed in the bottom left panel of Fig.~\ref{fig:erg-beta}, it is possible to find an input energy $E$ for which the maximal ergotropy of lower-temperature GMAD channels outperforms the results at higher temperatures; this effect is present also when considering the MAD ergotropy.
    
    \item For certain choices of the parameters, there is a state whose ergotropy is well preserved for any $E$ by the action of the channel, as reported in the bottom right panel of Fig.~\ref{fig:erg-beta}, where it also can be noted that in this case its presence is robust against temperature changes.

    \item Energy activation provides another counterintuitive effect of GMADs: despite being thermal operations, GMADs can “activate” passive input states, promoting their ergotropy to nonzero values. This collocates some GMADs outside of the set of the Passivity-Preserving Operations (PPO), the set of quantum channels preserving the passivity of the input \cite{singh2021partial}. An instance of this phenomenon is visible in the top left panel of Fig.~\ref{fig:erg-beta}, where the maximal ergotropy reaches non-vanishing values at $E=0$; i.e. for an input state coinciding with the ground state, that is passive.
\end{itemize}

Ball-maximal ergotropic functionals share most of these features; however, on the other end, they are always non-decreasing in input energy, since they are constructed by maximizing on energies up to $E$ the relative shell-maximal ergotropic functional, for each $E$.

\subsection{Ergotropic Capacitances}
\begin{figure}
    \centering
    \includegraphics[width=\linewidth]{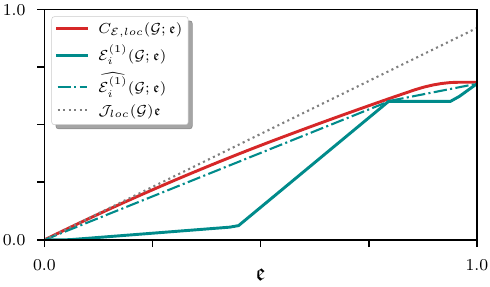}
    \caption{(Red solid curve) Local ergotropic capacitance for the GMAD channel $\G$ with $s_1=0.5$, $\bar{s}=0.745$, $\alpha_0=0.745$, $\beta=1$, acting on a three-level system with spectrum $(0,0.8,1)$. (Cyan solid curve) 1-cell ball-maximal incoherent ergotropy for the same channel; even if it is not concave in $\e$, its concave envelope (cyan dot-dashed curve), here indicated as $\widehat{\erg_i}$, bounds from below the respective capacitance. (Gray dotted line) Line with slope equal to the MAWER for the same channel; its value is $\mathcal{J}_{loc}(\G)\approx 0.919$ .}
    \label{fig:caps}
\end{figure}

As briefly introduced in Sec.~\ref{sec:ergfunc}, work capacitances assess the performance of the asymptotic use of the channel in a setting in which the energy per copy ratio $\e := E/n$ is kept fixed and finite.
Eventually, thanks to Eq. \eqref{eq:def_chi} and the concavity in $E$ of many of the ball-maximal ergotropy curves, it is possible to exactly evaluate work capacitances for many of the GMADs studied; an example is shown in Fig.~\ref{fig:caps}, where the local ergotropic capacitance is reported for a certain choice of parameters. Together with it is reported the 1-cell maximal incoherent ergotropy for the same GMAD, which is not concave in $\e$; in such cases, the concave envelope $\widehat{\erg_i}$ of the curve nevertheless provides a lower bound on the respective capacitance.

Another possibility in quantifying asymptotic performances of channels is by defining the Maximal Asymptotic Work-Energy Ratios (MAWERs); see Appendix \ref{app:workfunc} for a brief introduction or \cite{WEroleofnonlocalresource} for a thoroughly discussion. If the work capacitance is known for small values of $\e$, the respective MAWER can be found as the limit of the difference quotient for $\e\to0$ \cite{WEroleofnonlocalresource}; hence, for differentiable work capacitances, the MAWER can be found as the slope of the relative work capacitance at $\e=0$. The line whose slope is the MAWER is shown in gray in Fig.~\ref{fig:caps}.

\section{Coherent work dynamics under discrete evolution}
\label{sec:enact}
It is perhaps more interesting to consider what happens as one iterates the map $\G$ sequentially, since it allows us to investigate the behavior of ergotropy as a function of the number of iterations, which can be interpreted as a discretized time parameter. This approach provides a framework for investigating time-resolved effects, such as the {\it ergotropic Mpemba effect} (EME) \cite{medina2025anomalous, sapui2026ergotropic}, a non-monotonic behavior of the ergotropic loss of a noisy quantum battery in terms of the initial charge of the input states. As the EME has recently been observed for qubits and qutrits in \cite{sapui2026ergotropic}, we want to particularly stress on the relevance that the coherent part of the ergotropy retains in the economy of the effect: while it is certainly true that the incoherent ergotropy can play a dominant role in terms of magnitude of the observed trend inversion, the coherent EME is a genuinely quantum effect which has not a direct classical counterpart (in contrast to the incoherent EME that can be fully described by classical stochastic dynamics), as far as we restrict ourselves to finite size systems \cite{lu2017nonequilibrium, campisi2026unified}.

In the quest of observing if the GMAD can originate any instance of the coherent EME, we start by fixing a diagonal Hamiltonian $\hat{H}$ and we choose two initial states $\ro, \hat{\sigma}$ such that $\dephased{\rho} = \dephased{\sigma}$ (note that these states have the same fixed initial energy $E = \tr[\dephased{\rho}\, \hat{H}]$). With this choice, we can exploit the fact that for strictly incoherent operations the evolution of the level populations is detached from the one of the coherence to isolate the coherent contribution to the effect; in fact, since
\begin{equation}
    \G^n(\ro) = \G^n(\dephased{\rho}) + \G^n(\ro - \dephased{\rho})\quad \forall\,n\ ,
\end{equation}
we force the incoherent ergotropy of the two states $\ro, \hat{\sigma}$ to be the same for each $n$. Then, evaluating the ergotropy values at the output of the $n$-th iteration of $\G$, their difference only involves the coherent ergotropy changes. In Appendix \ref{app:cEMEtab} a pair of state is reported, for which we can observe coherent EME with a single application of the indicated GMAD.

\begin{figure}
    \centering
    \includegraphics[width=1\linewidth]{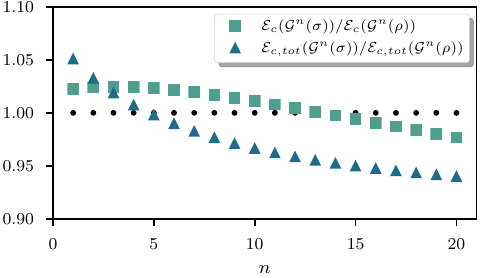}
    \caption{Ratios between the coherent ergotropy (green squares) and the total coherent ergotropy (blue triangles) of $\ro'$ and $\hat{\sigma}'$, for different map iterations $\G^n$. $\G$ is fixed by $s_1 = \sin(\frac{8}{10})$, $\bar{s} = \sin(\frac{\sqrt{5}}{10})$, $\alpha_0 = \frac{1}{\sqrt{5}}$, $\beta = \frac{1}{10}$.}
    \label{fig:coh-mpemba}
\end{figure}

\subsection{Total Coherent Ergotropy and Mpemba Effect}
An interesting fact emerges when one considers the availability of an asymptotically large number of identical copies of the state. As said in Sec.~\ref{sec:ergfunc}, in this framework, total ergotropy becomes the more appropriate performance quantifier for work extraction tasks.
In Eq. \eqref{eq:ergtot} the total ergotropy was introduced as a regularization of the ergotropy for the $n$-fold product state $\ro^{\otimes n}$; in the same way, we here introduce a regularization of the coherent ergotropy:
\begin{multline}
    \erg_{c, tot}(\ro;\hat{H}) \coloneqq \lim_{n\to+\infty}\frac{1}{n}\erg_{c}(\ro^{\otimes n};\hat{H}^{(n)}) = 
    \\
    =\lim_{n\to+\infty}\frac{1}{n} [\erg(\ro^{\otimes n};\hat{H}^{(n)})-\erg_{i}(\ro^{\otimes n};\hat{H}^{(n)})]\ .
\end{multline}
By fixing the incoherent basis of the $n$-fold Hilbert space as the tensor product of the local incoherent bases we have that $\Delta_{n\text{-fold}}(\ro^{\otimes n}) = (\Delta_{\text{local}}(\ro))^{\otimes n}$, and using Eq. \eqref{eq:erg_nc-is-erg-dephased} together with Eq. \eqref{eq:ergtot} we get
\begin{equation}
    \lim_{n\to+\infty}\frac{1}{n}\erg_{i}(\ro^{\otimes n}) = \erg_{tot}(\Delta_{\text{local}}(\ro)) = \erg_{tot}(\dephased{\rho})\ ;
\end{equation}
and thus,
\begin{equation}
    \erg_{c, tot}(\ro) = \erg_{tot}(\ro) - \erg_{tot}(\dephased{\rho}) = \tr[(\gibbs_{\delta_{\rho}}- \gibbs_{\rho})\hat{H}]\ .
\end{equation}
Therefore, any instance of the ergotropic Mpemba effect in this setting can be interpreted as follows: we observe the effect if, for some $n\in\mathbb{N}$, there are two states $\ro, \hat{\sigma}$, with $\dephased{\rho} = \dephased{\sigma}$, such that:
\begin{equation}
    \beta^\star_{\rho}>\beta^\star_{\sigma}\ \overset{\G^n}{\Longrightarrow}\ \beta^\star_{\rho'}<\beta^\star_{\sigma'}\ ,
\end{equation}
where $\ro' = \G^n(\ro)$, and similarly for $\hat{\sigma}'$. This recovers the usual formulation of the celebrated Mpemba effect and its generalizations to Markovian systems \cite{mpemba1969cool,lu2017nonequilibrium,ares2025quantum} in terms of the entropic inverse temperatures $\beta^\star$. In Fig.~\ref{fig:coh-mpemba}, the ratio between the coherent ergotropy of $\ro'$ and $\hat{\sigma}'$ is reported for different map iterations, along with the same ratio for the total coherent ergotropy; crossing the line with ordinate $y=1$ corresponds to a flip in the ordering of the quantities.

\section{Conclusions}
\label{sec:concl}
We introduced the GMAD channel as a novel formal approach to describe noise and decoherence in generic qudit systems at arbitrary environmental temperatures.
Our approach generalizes the MAD to the finite-temperature regime, as well as the GADC beyond the $d = 2$ case.
To investigate its properties from the perspective of work extraction, we introduced new functionals, such as the Maximally Coherent and Maximally Incoherent Work Functionals, and performed an in-depth analysis of a paradigmatic qutrit model.
Our results reveal counterintuitive features of the ergotropic work functionals, showing that they are not necessarily monotonic in temperature: low-temperature channels can, in some cases, exhibit better thermodynamic performance than high-temperature ones.
The introduction of the GMADs and of these new work functionals opens several directions for future research, including investigating the possible advantage of using entangled initial states. Very recently, optimal performances of quantum engines with unitary strokes have been studied \cite{vs2026ergotropy}; GMAD could help model a richer thermalization dynamics in such scenarios. Another interesting direction would be a more in-depth analysis of the relation between the dynamical resource theories of thermodynamics \cite{chiribella2021fundamental} and the quantum capacitances introduced in \cite{QWC} and in this manuscript.\\

\section*{Acknowledgments} We thank Stefano Chessa for useful discussions. We acknowledge financial support by MUR (Ministero dell’Universit{\`a} e della Ricerca) through the PNRR MUR project PE0000023-NQSTI.

\appendix

\begin{widetext}
\section{Microscopic derivation of the GMAD}
\label{app:microGMAD}

GMADs model the contact of a system with a thermal environment, extrapolating the MAD channel to the finite temperature case.
A common model to represent the dissipation induced by a thermal environment is to use a thermal operation, that is, an energy-preserving interaction of a system with an environment $\mathcal{H}_B$ prepared in an equilibrium state $\gibbs_B = \frac{e^{- \beta \hat{H}_B}}{Z}$,
where we assumed that the Hamiltonian of the environment is $\hat{H}_B$ and $Z \coloneqq Tr[e^{- \beta \hat{H}_B}]$. Denoting the global system-environment evolution by $\hat{U}$, we have
\begin{equation} \label{eq:thermop}
    [\hat{U}, \hat{H}_S + \hat{H}_B] =0,
\end{equation}
where we also added a subscript $S$ to the system Hamiltonian $\hat{H}_S=\sum_{j}\epsilon_j\proj{j}$, for the sake of clarity. The condition \eqref{eq:thermop} implies that the unitary evolution is the direct sum of blocks $\hat{U}^{(E)}$ acting on subspaces of total energy $E$:
\begin{equation} \label{eq:blockun}
    \hat{U} = \bigoplus_{E} \hat{U}^{(E)}, 
\end{equation}
where $\hat{U}^{(E)}$ acts in the subspace $\mathcal{H}^{(E)}_{SB} \in \mathcal{H}_S \otimes \mathcal{H}_B$, characterized by the condition 
\begin{equation} \label{eq:UEE}
    (\hat{H}_S + \hat{H}_B) \ket{\psi} = E \ket{\psi}, \qquad \forall \, \ket{\psi} \in  \mathcal{H}^{(E)}_{SB}.
\end{equation}
From the definition given above, it is easy to show that a thermal operation admits a representation in terms of the following off-diagonal Kraus operators
\begin{equation} \label{eq:krausgen} \hat{K}_{\mu \nu} = \sqrt{\tau_{\nu} } \bra{\mu} \hat{U} \ket{\nu},
\end{equation}
where $\ket{\mu}, \ket{\nu}$ are eigenstates of $\hat{H}_B$ with energies $\varepsilon_{\mu},\varepsilon_{\nu}$ and $\tau_{\nu} \coloneqq \bra{\nu} \gibbs_B \ket{\nu}$.
Notice that, due to the energy preservation requirement in Eq.~\eqref{eq:thermop}, if we define $\Delta_{\mu \nu} \coloneqq \varepsilon_{\mu} - \varepsilon_{\nu}$ and $\omega_{ij} \coloneqq \epsilon_i - \epsilon_j$ we also have
\begin{equation}
    \hat{K}_{\mu \nu} = \sum_{i,j} \kappa_{ij} \delta(\omega_{ij}
    -\Delta_{\nu \mu}) \ket{i}\bra{j} 
\end{equation}
for some coefficients $\kappa_{ij}$, that is, the Kraus operator corresponding to the environmental jump between $\nu$ and $\mu$ can only induce jumps with energy gap equal to $- \Delta_{\mu \nu} $ in the system.
We are interested in thermal operations that reduce to the MAD when the temperature of the initial state of the environment is $0$, that is, in the limit $\beta \rightarrow \infty$.
In this regime, the only $\nu$ for which $\tau_{\nu}$ does not vanish is $\nu=0$, associated with the ground state of the environment, that we assume to be unique.
The only off-diagonal Kraus operators that survive in this limit, in which $\tau_{0} \rightarrow 1$ are 
\begin{equation}
  \hat{K}_{\mu 0} = \bra{\mu} \hat{U} \ket{0}  = 
  \sum_{i,j} \kappa_{ij} \delta(\omega_{ij}
    -\Delta_{0 \mu}) \ket{i}\bra{j},
\end{equation}
for all $\mu \neq 0$.
To recover the most general class of MADs, we need every system jump $\ketbra{i}{j}$ to be present at least once in the representation above.
This means that for every possible gap $\omega_{ij} >0$ there should be an environmental state $\ket{\mu}$ with energy $\Delta_{\mu 0} = \omega_{ij}$.
Assuming (for simplicity) that $\epsilon_{i} \neq \epsilon_{j}$ for $i \neq j$ and $\omega_{ij} \neq \omega_{nm}$ for $(i,j) \neq (n,m)$, there must be at least one energetic level of the environment associated with every gap in the system; the easiest choice is to create a bijective correspondence between gaps of the system $\omega_{ij}$ and the excited levels of the environment, so that we can index the environment energy eigenstates as $\ket{\nu}\equiv\ket{\omega_{ij}}$.
For a system with dimension $d$, there are $d(d-1)/2$ positive energy gaps (all the possible unordered pairs of levels).
We conclude that the simplest thermal operation reducing to the MAD in the zero-temperature limit is realized by coupling the system to a $d(d-1)/2 +1$-dimensional environment with a ground state $\ket{0}$ and a collection of excited states $\ket{\omega_{ij}}$, each one with associated energy $\omega_{ij}$.
This structure is summarized in Fig.~\ref{fig:sketch2}.
\begin{figure}
    \centering
    \includegraphics[width=0.7\linewidth]{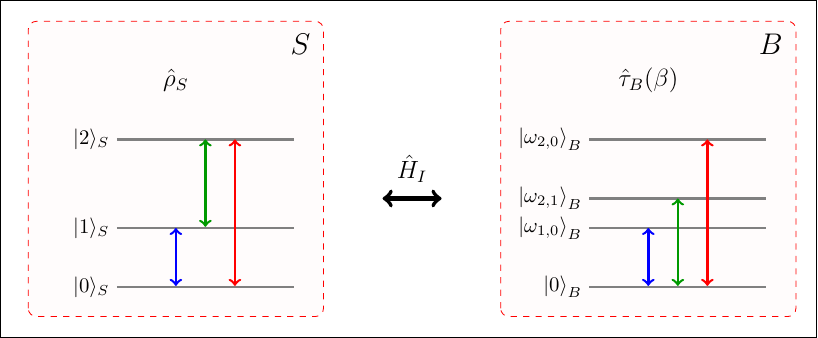}
    \caption{Depiction of the level structure of the environment $B$, and its dependence on the system spectrum, for the case of a qutrit system $S$.}
    \label{fig:sketch2}
\end{figure}

To complete the characterization, we have to discuss the form of the unitary $\hat{U}$ acting on the system-environment compound. For this sake, it is sufficient to describe the action of each one of the unitaries $\hat{U}^{(E)}$ defined in Eq. \eqref{eq:blockun}.
According to the discussion above, it should be possible to describe all the transitions coupling $\ket{i} \otimes \ket{0}$ with $\ket{j} \otimes \ket{\omega_{ij}}$, for arbitrary $i,j \in [0,\ldots, d-1]$ in the currently introduced GMAD formalism. Since these transitions happen within subspaces with total energy $\epsilon_i$, the unitary blocks $\hat{U}^{(\epsilon_i)}$ for some $i$ will be assumed to be generic unitary matrices, their choice eventually influencing the form of the coefficients of Eqs. \eqref{eq:offkraus}, \eqref{eq:diagkraus}.
On the contrary, we will assume
\begin{equation}
    \hat{U}^{(E)} = \mathds{I} \ \  {\rm if} \ \ \nexists\  i \in [0, \ldots, d-1] \  {\rm such \, that } \  E = \epsilon_i.
\end{equation}
Below we prove that the choices made above result in a Kraus representation corresponding to Eqs. \eqref{eq:offkraus} and \eqref{eq:diagkraus} in the main text.

\subsection{Microscopic derivation of the Kraus operators}
The starting point to associate a general structure to the microscopic model outlined in Fig.~\ref{fig:sketch2} is to write the spaces $\mathcal{H}_{SB}^{(E)}$ introduced in Eq. \eqref{eq:UEE}.
We can divide these subspaces in two categories:
\begin{enumerate}
    \item {\it Non-trivial fixed energy subspaces.} In these subspaces the energy $E$ is exactly equal to $\epsilon_i$ for $i\in [0,d-1]$, that is, to one of the eigenvalues of the system Hamiltonian. 
    \item  {\it Trivial fixed energy subspaces.} All the subspaces $\mathcal{H}^{(E)}_{SB}$ in which the condition above is not satisfied.
\end{enumerate}
The rationale behind such a distinction is that the system-environment block unitary evolution $\hat{U}^{(E)}$ acts like the identity inside the trivial energy subspaces. 
We will thus focus on non-trivial energy subspaces, that are in number equal to $d$.
The non-trivial space $\mathcal{H}^{(\epsilon_k)}_{SB}$ comprises the vectors:
\begin{equation} \label{eq:list}
    \{\ket{k,0},\ket{k-1,\omega_{k,k-1}},\ket{k-2,\omega_{k,k-2}},\cdots,\ket{r,\omega_{k,r}},\cdots,\ket{k-1,\omega_{k,k-1}},\ket{0,\omega_{k,0}}\}\ ;
\end{equation}
thus, it is $(k+1)$-dimensional.
The unitary $\hat{U}^{(\epsilon_k)}$ associated to this block takes the form:
\\
\begin{table}[H]
\centering
\begin{tabular}{c|c c c c c c} 
                                & $\ket{k,0}$ & $\ket{k-1,\omega_{k,k-1}}$ & $\cdots$ & $\ket{r,\omega_{k,r}}$ & $\cdots$ & $\ket{0,\omega_{k,0}}$\\
                                \hline
                                
                    $\bra{k,0}$ & $u_{k,k}^{(k)}$ & $u_{k,k-1}^{(k)}$ & $\cdots$ & $u_{k,r}^{(k)}$ & $\cdots$ & $u_{k,0}^{(k)}$\\
     $\bra{k-1,\omega_{k,k-1}}$ &$u_{k-1,k}^{(k)}$&$u_{k-1,k-1}^{(k)}$& $\cdots$ &$u_{k-1,r}^{(k)}$& $\cdots$ &$u_{k-1,0}^{(k)}$\\
                       $\vdots$ & $\vdots$          & $\vdots$            &          & $\vdots$          &          & $\vdots$ \\
         $\bra{r,\omega_{k,r}}$ & $u_{r,k}^{(k)}$ & $u_{r,k-1}^{(k)}$ & $\cdots$ & $u_{r,r}^{(k)}$ & $\cdots$ & $u_{r,0}^{(k)}$ \\
                       $\vdots$ & $\vdots$          & $\vdots$            &          & $\vdots$          &          & $\vdots$ \\
         $\bra{0,\omega_{k,0}}$ & $u_{0,k}^{(k)}$ & $u_{0,k-1}^{(k)}$ & $\cdots$ & $u_{0,r}^{(k)}$ & $\cdots$ & $u_{0,0}^{(k)}$\\
\end{tabular}
\caption{The $\hat{U}^{(\epsilon_k)}$ block.}
\label{tab:table}
\end{table}
\noindent where we used the superscript $(k)$ instead of $(\epsilon_k)$ for ease of notation.
It is important to note that, given a certain frequency $\omega_{ij}$, the ket $\ket{\ldots, \omega_{ij}}$ only appears in a single non-trivial energy subspace.
In fact, suppose there are two different subspaces $\mathcal{H}^{(\epsilon_k)}_{SB}$, $\mathcal{H}^{(\epsilon_{k'})}_{SB}$ for $k \neq k'$ such that the frequency $\omega_{kr}$ belongs to both, i.e. there exist system states $\ket{r},\ket{r'}$ such that $\ket{r,\omega_{kr}}$ belongs to the block $k$ and $\ket{r',\omega_{kr}}$ belongs to $k'$. This would imply that 
\begin{equation}
    \epsilon_{k'} = \epsilon_{r'}+\omega_{kr} = \epsilon_{r'}+\epsilon_k - \epsilon_r \quad\text{with}\quad r'\neq r,
\end{equation}
and it would violate the assumption that the energy gaps in the system are non-degenerate.
We can thus associate every excited state of the bath $\ket{\omega_{kr}}$ to its non-trivial unitary, in this case $\hat{U}^{(\epsilon_k)}$.
The only exception to this behavior is the ground state of the environment, since the vector $\ket{\epsilon_k, 0}$ appears in the subspace with energy $\epsilon_k$ for all $k \in [0, d-1]$ [see Eq. \eqref{eq:list}].
\\
We are now ready to compute the Kraus operators, using the definition in Eq. \eqref{eq:krausgen} and the form of the unitary blocks $\hat{U}^{(\epsilon_k)}$ identified above.
The off-diagonal Kraus operators can be obtained by projecting on environmental energy levels $\ket{\omega_{ij}}, \ket {\omega_{nm}}$.
The off-diagonal Kraus operators thus read
\begin{equation}
    \hat{K}_{\omega_{ij}\,,\, \omega_{nm}}
    =    \sqrt{\tau_{nm}} \bra{\omega_{ij}} \hat{U} \ket{\omega_{nm}}
\end{equation}
where 
\begin{equation}
    \tau_{ab}:=\frac{e^{-\beta \omega_{ab}}}{Z}=\frac{e^{-\beta (\epsilon_a - \epsilon_b)}}{Z}\ , \quad    Z \coloneqq \sum_{a > b} e^{- \beta \omega_{ab}} + 1\ ,
\end{equation}
where the value $1$ above corresponds to the contribution of the ground level of the environment. Here $\tau_{ab}$ does not refer to the system Gibbs state $\gibbs$, as it should be clear by the context.

Of the Kraus operators introduced above, most are actually equal to $0$.
Indeed, since the system-environment interaction is energy preserving, non-vanishing contributions must satisfy
\begin{equation}
    \exists \: k,h,l \, \, {\rm such \, that} \quad 
    \omega_{ij} + \epsilon_k = \omega_{nm} + \epsilon_h = \epsilon_l \ ,
\end{equation}
that is
\begin{align} \label{eq:conds}
    \omega_{ij} =  \epsilon_l - \epsilon_k \ ,\qquad
    \omega_{nm} =  \epsilon_l -  \epsilon_h \ ;
\end{align}
but due to the non-degeneracy of the system energy gaps, we must have $l = i = n$, $k = j$ and $h = m$.
This means that 
\begin{equation}
    \hat{K}_{\omega_{ij}\,,\, \omega_{nm}}
    =  \delta_{in} \sqrt{\tau_{nm}}\bra{\omega_{ij}} \hat{U}^{(\epsilon_i)} \ket{\omega_{nm}}\,.
\end{equation}
Remembering the condition of energy preservation we have (see also the table \ref{tab:table})
\begin{equation}
  \bra{\omega_{ij}} \hat{U}^{(\epsilon_i)} \ket{\omega_{im}} = u_{jm}^{(i)} \ket{j}\bra{m}\ ,
\end{equation}
where by unitarity of $\hat{U}^{(\epsilon_i)}$ we have $ \sum_{j} u_{jm}^{(i)} u_{jm'}^{(i)\,*} = \delta_{mm'} $.
Selecting only non-zero Kraus operators with $i=n$, we obtain
\begin{equation}
   \hat{K}_{\omega_{ij}\,,\, \omega_{im}} 
    =    u_{jm}^{(i)}  \sqrt{\tau_{i m}} \ket{j}\bra{m}\ .  
\end{equation}
These Kraus operators are identified by $3$ indices. While $j$ and $m$ are different and free to take all the values between $0$ and $d-1$, the index $i$ is constrained to be greater or equal to $j$ and $m$, see condition \eqref{eq:conds}. 
For each $i$ from 0 to $d-1$, the two indices $j,m$ take values from 0 to $i$ included, but $j\neq m$; so, the total number of off-diagonal Kraus operators is:
\begin{equation}
    \sum_{i=1}^{d}(i^2-i) = \frac{d(d^2-1)}{3}\ . 
\end{equation}
\\
Now we turn our attention to diagonal Kraus operators, starting from
\begin{equation}
    \hat{K}_{00} \coloneqq {Z}^{-\frac{1}{2}}\bra{0} \hat{U} \ket{0}.
\end{equation}
Given our choice of the environment, no states of the form $\ket{\ldots,0}$ appear in trivial subspaces, since the total energy of any state of the system paired with the ground state of the environment is exactly equal to one of the $\epsilon_j$, falling in the non-trivial block of energy $\epsilon_j$. 
Using the form of the unitary in table \ref{tab:table}, it is immediate to derive
\begin{equation}
    \hat{K}_{00} = {Z}^{-\frac{1}{2}}\bigg(\ketbra{0}{0}+\sum_{k=1}^{d-1} u_{kk}^{(k)}\ketbra{k}{k}\bigg) = {Z}^{-\frac{1}{2}}\bigg(\sum_{k=0}^{d-1} u_{kk}^{(k)}\ketbra{k}{k}\bigg) \quad\text{ with }\quad u_{00}^{(0)}=1.
\end{equation}

We are left with evaluating the remaining diagonal Kraus operators; the process to obtain them should be identical to the one for $\hat{K}_0$, but we have to pinpoint all the trivial spaces.
\\
We see that in each (k+1)-dimensional block we have $u_{rr}^{(k)}\ketbra{r}{r}$ if we project over $\ket{r,\omega_{kr}}$. We have no other contributions from the nontrivial blocks, since the frequency ket $\ket{\omega_{kr}}$ appears in that subspace only.
So, we just miss the contributions from the trivial blocks. These have to be the cases in which the system is in a state from where transitions at $\omega_{kr}$ would be off-resonance.
Since, from our hypotheses, for a fixed bath ket $\ket{\omega_{ij}}$ we only have one resonant transition (if we had two we would again introduce degeneracy), we can find the trivial fixed energy subspaces as the ones like $\{\ket{s,\omega_{ij}}\}$ with $s\neq j$. 
\\
For example, in the qutrit case they are: $\{\ket{0,\omega_{21}}\}$ and $\{\ket{2,\omega_{21}}\}$ for $\omega_{ij}\equiv\omega_{21}$; $\{\ket{2,\omega_{20}}\}$, $\{\ket{1,\omega_{20}}\}$ for $\omega_{ij}\equiv\omega_{20}$; $\{\ket{2,\omega_{10}}\}$, $\{\ket{1,\omega_{10}}\}$ for $\omega_{ij}\equiv\omega_{10}$. 
Hence, adding all up:
\begin{equation}
    \hat{K}_{\omega_{ij}\,,\, \omega_{ij}} = \sqrt{\tau_{ij}}\bigg(\underbrace{\sum_{l\neq j}\ketbra{l}{l}}_{\text{trivial blocks}} + \underbrace{u_{jj}^{(i)}\ketbra{j}{j}}_{\text{nontrivial block}}\bigg) = \sqrt{\tau_{ij}}\Big(\identity+(u_{jj}^{(i)}-1)\ketbra{j}{j}\Big).
\end{equation}
The number of these diagonal Kraus operators is $d(d-1)/2$.
We can now relabel the indices in the Kraus operators to obtain the notation used in the main text.
We introduce an index $m = 0,1,..., d-1$, as follows
\begin{equation}
    \hat{K}^{(m)}_{ji} \coloneqq  \hat{K}_{\omega_{mi}\,,\, \omega_{mj}} \ \  {\rm for} \,\, m \neq 0\,,\ i \neq j\,; \qquad  \hat{K}^{(D)}_{ij} \coloneqq 
    \hat{K}_{\omega_{ij}\,,\, \omega_{ij}}\ \  {\rm for}\ i > j\,; \qquad \hat{K}^{(D)}_{00} \coloneqq \hat{K}_0\,;
\end{equation}
with $i,j \in [0,d-1]$.
The output state of the map can be explicitly constructed by using
\begin{equation}
    \label{eq:outDdim}
    \begin{aligned}
        \G(\ketbra{p}{q}) &= \delta_{pq}\bigg(\sum_{m=\max(1,\,p)}^{d-1}\sum_{\substack{i=0 \\ i\neq p}}^{m}|\g_{ip}^{(m)}|^2\ketbra{i}{i}\bigg)\ +
        \\
        &+\bigg(\frac{1}{Z}u_{pp}^{(p)}u_{qq}^{*\,(q)} + \sum_{l=1}^{d-1}\sum_{j=0}^{l-1}\tau_{lj}(\bar\delta_{jp}\bar\delta_{jq} + \delta_{jp}\bar\delta_{jq}u_{pp}^{(l)} + \bar\delta_{jp}\delta_{jq}u_{qq}^{*\,(l)} + \delta_{jp}\delta_{jq}|u_{pp}^{(l)}|^2)\bigg)\ketbra{p}{q}\ ,
    \end{aligned}
\end{equation}
where $\bar\delta_{ab} \coloneqq 1 - \delta_{ab}$. The quantity
\begin{equation}
    \gamma_{pp}\coloneqq\bigg(\frac{1}{Z}|u_{pp}^{(p)}|^2 + \sum_{l=1}^{d-1}\sum_{j=0}^{l-1}\tau_{lj}(\bar\delta_{jp} + \delta_{jp}|u_{pp}^{(l)}|^2)\bigg)
\end{equation}
represents the \textit{survival rate} of the level $p$.
In the case of a three-level system, the output state becomes:
\begin{equation}
    \begin{aligned}
        &\G(\ro) = 
        \\
        &\!\!\!\!\begin{bmatrix}
            \gamma_{00}\rho_{00} + (|\g_{01}^{(1)}|^2+ |\g_{01}^{(2)}|^2) \rho_{11} + |\g_{02}^{(2)}|^2\rho_{22} & \alpha_{01}\rho_{01} & \alpha_{02}\rho_{02}
            \\
            & \gamma_{11}\rho_{11} + (|\g_{10}^{(1)}|^2  + |\g_{10}^{(2)}|^2)\rho_{00} + |\g_{12}^{(2)}|^2\rho_{22} & \alpha_{12}\rho_{12}
            \\
            & & \gamma_{22}\rho_{22} + |\g_{20}^{(2)}|^2\rho_{00} + |\g_{21}^{(2)}|^2\rho_{11}
        \end{bmatrix}
    \end{aligned}
\end{equation}
with
\begin{equation}
    \begin{aligned}
        &\gamma_{00} \coloneqq (Z^{-1} + \tau_{10}|u_{00}^{(1)}|^2 + \tau_{20}|u_{00}^{(2)}|^2+ \tau_{21})\ ,
        \\
        &\gamma_{11}\coloneqq (Z^{-1}|u_{11}^{(1)}|^2 + \tau_{10} + \tau_{20} + \tau_{21}|u_{11}^{(2)}|^2)\ ,
        \\
        &\gamma_{22}\coloneqq (Z^{-1}|u_{22}^{(2)}|^2 + \tau_{10} + \tau_{20} + \tau_{21} )\ ;
        \\
        &\alpha_{01} \coloneqq (Z^{-1}u_{11}^{*\,(1)} + \tau_{10}u_{00}^{(1)} + \tau_{20}u_{00}^{(2)} + \tau_{21}u_{11}^{*\,(2)})\ ,
        \\
        &\alpha_{02} \coloneqq (Z^{-1}u_{22}^{*\,(2)} + \tau_{10}u_{00}^{(1)} + \tau_{20}u_{00}^{(2)} + \tau_{21})\ ,
        \\
        &\alpha_{12} \coloneqq (Z^{-1}u_{11}^{(1)}u_{22}^{*\,(2)} + \tau_{10} + \tau_{20} + \tau_{21}u_{11}^{(2)})\ .
    \end{aligned}
\end{equation}
\section{Minimal Set of Kraus Operators for the GMAD}
\label{app:minimalset}
In the main text we defined the GMAD using a representation in terms of a thermal operation.
This representation allows us to make the physical mechanism lying behind the GMAD very clear, as a thermalizing operation the preserves the total energy of the system and its environment.
However, it is clear that the Kraus operators obtained in this way do not constitute a minimal Kraus decomposition, due to their excessive number.
In this section we find a minimal Kraus decomposition for every GMAD channel.
As a first step, we show that it is possible to reduce the set of the $d(d^2-1)/3$ off-diagonal Kraus operators to a set of $d(d-1)/2$ 
operators.
This is simply done by considering the Kraus operators inducing the same jumps on the system and merging them in a single Kraus operator.
We have
\begin{equation}
    \sum_{m = \max{ i,j}}^{d-1}\gamma_{ij}^{(m)} \gamma_{ij}^{(m)*}\ket{i}\bra{j} \ro \ket{j}\bra{i} = \eta_{ij}  \ket{i}\bra{j} \ro \ket{j}\bra{i}
\end{equation}
where $\eta_{ij} =  \sum_{m \geq i,j} \gamma_{ij}^{(m)}  \gamma_{ij}^{(m)\,*}$.
We can thus define a minimal set of off-diagonal Kraus operators as
\begin{equation}
    \tilde{K}_{i,j} = \sqrt{\eta_{ij}}\ket{i}\bra{j}\,,\ \text{for}\ i<d\ ,\,j<d\,,\,i\neq j\ .
\end{equation}
The situation is slightly more involved for the diagonal Kraus operators.
The diagonal Kraus operators form a non-trace preserving CP map given by
\begin{equation}
    \Phi_D(\ro) = \sum_{s=0}^{\frac{d(d-1)}{2}} \hat{K}^{(D)}_{s} \ro \hat{K}^{(D)\dag}_{s}.
\end{equation}
where $s$ is an index running on all the ordered pairs: $s\in \{0\}\cup\{(ij)\,,\ i>j\}$. We start by decomposing the Kraus operators in the diagonal projectors
\begin{equation}
 \Phi_D(\ro) = \sum_{l,h=0}^{d-1} \sum_{s}^{\frac{d(d-1)}{2}} c^{(s)}_{h} c^{(s) *}_{l}  \ketbra{h}{h}\,  \ro\,\ketbra{l}{l} .   
\end{equation}
We can now define the positive semidefinite matrix
\begin{equation}
    Q_{hl} = \sum_{s}^{\frac{d(d-1)}{2}}  c^{(s)}_{h} c^{(s) *}_{l},
\end{equation}
and diagonalize it
\begin{equation}
     Q_{hl}  = \sum_r {V}_{hr} D_r  V_{lr}^*\ .
\end{equation}
Eventually, we define a new set of $d$ Kraus operators 
\begin{equation}
    \tilde{K}^{(D)}_r = \sqrt{D_r}  \sum_h V_{hr} \ketbra{h}{h}\,,\ \text{for}\ r<d\ ,
\end{equation}
from which we get:
\begin{equation}
    \Phi_D(\ro) = \sum_{r=0}^{d-1}  \tilde{K}^{(D)}_r \ro  \tilde{K}^{(D)\dag}_r \ .
\end{equation}

\section{Properties of Maximal Coherent and Incoherent Work Functionals}
\label{app:Workcoh}
Let us start by defining the \textit{totally dephasing channel} $\Delta(\cdot)$. Given a Hamiltonian $\hat{H} = \sum_{l} \epsilon_l \proj{l}$, we define $\Delta(\cdot)$ as
\begin{equation}
    \label{eq:defDelta}
    \Delta(\ro) = \sum_{l=0}^{d-1} p_l \proj{l}\ ,
\end{equation}
with $p_l \coloneqq \bra{l}\ro\ket{l}$. The channel $\Delta(\cdot)$ maps every state $\ro$ to the diagonal state $\dephased{\rho} \coloneqq\Delta(\ro)$.
The incoherent ergotropy \eqref{eq:def-erg_nc}, which for reader's commodity we redefine here,
\begin{equation}
    \label{eq:def-erg_i-app}
    \erg_{i}(\ro; \hat{H}) := \max_{\hat{V}\in \mathbb{U}_{i}(d)}\left\{\tr[\ro \hat{H}]-\tr[\hat{V}\ro \hc{\hat{V}} \hat{H}]\right\} = \tr[\ro \hat{H}]-\min_{\hat{V}\in \mathbb{U}_{i}(d)}\tr[\hat{V}\ro \hc{\hat{V}} \hat{H}]\ ,
\end{equation}
directly inherits some properties from ergotropy, also thanks to the relation $\erg_{i}(\ro) = \erg(\dephased{\rho})$. In particular, $\erg_{i}$ is continuous (as it is a composition of continuous functions) and convex in $\ro$:
\begin{multline}
    \erg_{i}(q\ro_1 + (1-q)\ro_2) = \erg(\Delta(q\ro_1 + (1-q)\ro_2))= 
    \erg(\Delta(q\ro_1) + \Delta((1-q)\ro_2)) \leq
    \\
    \leq \erg(\Delta(q\ro_1)) + \erg((1-q)\Delta(\ro_2)) = q\erg_{i}(\ro_1) + (1-q)\erg_{i}(\ro_2)\ .
\end{multline}
Incoherent ergotropy is also \textit{super-additive} for factorized independent systems, that is:
\begin{equation}
    \erg_{i}(\ro_1\otimes\ro_2\,;\hat{H}_1 + \hat{H}_2) \geq \erg_{i}(\ro_1\,; \hat{H}_1) + \erg_{i}(\ro_2\,; \hat{H}_2)\ .
\end{equation}
To prove this, let us explicitly rewrite $\erg_{i}(\ro_1\otimes\ro_2)$ as $\erg(\dephased{\rho_1\otimes\rho_2})$. In this case, we are considering the ergotropy of the state of two systems, dephased in the eigenbasis of the collective Hamiltonian $\hat{H}_{12} = \hat{H}_1 + \hat{H}_2$. Therefore, the energy of the passive state of $\dephased{\rho_1\otimes\rho_2}$ is always lower or equal to the sum of the energies of the passive states for $\dephased{\rho_1}$ and $\dephased{\rho_2}$, since the possibility of applying global permutations in general helps reaching lower passive states' energies \cite{alimuddin2019bound}.

The $n$-cells maximal incoherent ergotropy is
\begin{equation}
    \erg_{i}^{(n)}(\Lambda;E) \coloneqq \max_{\ro^{(n)}\,\in\,\mathfrak{I}^{(n)}_E}\ {\erg}_{i}(\Lambda^{\otimes n}(\ro^{(n)});\hat{H}^{(n)})
\end{equation}
Since $\erg_i(\ro)$ is continuous in $\ro$, the maximal incoherent ergotropy is continuous in $E$, thanks to Berge's maximum theorem \cite{berge1963espaces}. The incoherent ergotropic work capacitance is defined by taking the limit of the supremum as $n$ tends to $\infty$: as in \eqref{eq:def-capacity},
\begin{equation}
    C_{\erg_i}(\Lambda;\e) \coloneqq \limsup_{n \to \infty}\frac{\erg_i^{(n)}(\Lambda;E=n\e)}{n}\ .
\end{equation}
The super-additivity for incoherent ergotropy allows us to state the existence of the proper limit in \eqref{eq:def-capacity} (without the $\sup$) by following the same line of reasoning that was used in \cite{QWC}.

The coherent ergotropy $\erg_{c}(\cdot) = (\erg - \erg_{i})(\cdot)$ is continuous in its arguments, but lacks definite convexity properties; furthermore, we present here a simple example to show that $\erg_{c}$ is not even \textit{quasiconvex}, i.e.
\begin{equation}
    \erg_{c}(p\ro_1+(1-p)\ro_2) \not\leq \max\{\erg_{c}(\ro_1),\erg_{c}(\ro_2)\}\ .
\end{equation}
Quasiconvexity is a much weaker condition than convexity; however, it suffices to take two isoenergetic states of the form
\begin{equation}
    \ro_1=\begin{pmatrix}
        \frac{1}{3} -\lambda & 0 & \frac{1}{3} -\lambda\\
         & \frac{1}{3} + 2\lambda & 0\\
          &   & \frac{1}{3} -\lambda
    \end{pmatrix}\ ,\quad
    \ro_2=\begin{pmatrix}
        \frac{1}{3} +\lambda &0 & \frac{1}{3} +\lambda\\
          & \frac{1}{3} -2\lambda & 0\\
          &   & \frac{1}{3} +\lambda
    \end{pmatrix}\ ,
\end{equation}
for $\lambda\in[0,1/6]$, to see that the mixture $(\ro_1+\ro_2)/2$ has higher coherent ergotropy than each of the states. Note anyway that for any partial dephasing $\Delta_p(\ro) \coloneqq (1-p)\ro + p\Delta(\ro)$ one has
\begin{equation}
    \begin{aligned}
        \erg_c(\ro)-\erg_c(\Delta_p(\ro)) &= \erg(\ro) - \erg_i(\ro) - \erg(\Delta_p(\ro)) + \erg_i(\Delta_p(\ro))=
        \\
        &=\tr[(\ro-\pass{\rho})\hat{H}] - \tr[(\Delta_p(\ro)- \pass{\Delta_p(\rho)})\hat{H}]=
        \\
        &=\tr[\pass{\Delta_p(\rho)}\hat{H}] - \tr[\pass{\rho}\hat{H}]\geq 0\ .
    \end{aligned}
\end{equation}
The last inequality comes from the fact that $\Delta_p$ is unital and the hypotheses of the Uhlmann's majorization theorem hold (known also as Alberti––Uhlmann's theorem, reported in \cite{alberti1982stochasticity}; see also \cite{hiai1987majorization}); the Schur–concavity of the passive state's energy eventually leads to the inequality. Nevertheless, the optimal state for coherent ergotropy is pure. To prove this, given any non-pure state $\ro$, consider the pure state $\hat\psi_\rho \coloneqq \proj{\psi_\rho}$with the same populations (and thus energy) of $\ro$:
\begin{equation}
    \ket{\psi_\rho} = \sum_j \sqrt{p_j}e^{i\theta_j}\ket{j}\ ;\quad \dephased{\rho} = \dephased{\psi_\rho}\ .
\end{equation}
Evaluating the coherent ergotropy on both, and subtracting, we have
\begin{equation}
        \erg_c(\hat\psi_\rho)-\erg_c(\ro) =\erg(\hat\psi_\rho) -\erg(\ro) = \tr[\pass{\rho}\hat{H}]\geq 0\ ,
\end{equation}
since for pure states energy and ergotropy coincide (setting the ground state energy to zero).

The maximal coherent ergotropy is continuous in $E$ (again for Berge's theorem), but it only permits a much less detailed characterization than the incoherent one. Although the properties of coherent ergotropy just shown preclude any statement on the features of the optimal input state in general, something can definitely be said for a certain kind of maps. For the GMAD in particular, consider again the input states $\ro, \hat{\psi}_\rho$, which are related by
\begin{equation}
    \rho_{ii} = {\psi}_{ii}\,,\ \ \forall\,i\,;\quad \rho_{ij} = \eta_{ij}\psi_{ij}\,,\ \ \forall\,i,j\,\ \ (i\neq j);
\end{equation}
where $\eta_{ij}$ are constants in $[0,1]$.
By defining $\eta_{ii} \coloneqq 1\ \forall i$, we can rewrite this as $\ro = \mathcal{S}_\eta(\hat{\psi}_\rho) \coloneqq \hat\eta \circ \hat{\psi}_\rho$, with ``$\circ$" the Schur product.
The matrix $\hat\eta$ is positive-semidefinite (PSD):
\begin{equation}
    \eta_{ij} = \frac{\rho_{ij}}{\psi_{ij}} = \left(\frac{1}{\sqrt{p_i}e^{i\theta_i}}\right)\rho_{ij}\left(\frac{1}{\sqrt{p_j}e^{-i\theta_j}}\right) =\colon D_{ii}{\rho_{ij}}D_{jj}^* = [\hat{D}{\ro}\hat{D}^\dagger]_{ij}
\end{equation}
where $\hat{D}$ is a diagonal matrix with $D_{ii}\coloneqq(\sqrt{p_i}e^{i\theta_i})^{-1}$ (assuming finite populations; if this condition is not met, it suffices to restrict ourselves to the support of $\ro$). Thus, $\hat{\eta}$ inherits the PSD property from $\ro$ via Sylvester’s law of inertia \cite{bhatia2009positive}.
Now consider the action of the map $\G$. As visible in Eq. \eqref{eq:outDdim}, it too acts on the off-diagonal terms as a Schur multiplier; let us call it $\hat{\zeta}$. For the output states $\G(\ro)$ and $\G(\hat{\psi}_\rho)$ we have that $\Delta(\G(\ro)) = \Delta(\G(\hat{\psi}_\rho))$ for the SIO property, while the action of the map on off-diagonal terms reads 
\begin{equation}
    [\G(\ro)]_{ij} = \zeta_{ij}\rho_{ij} = \zeta_{ij}\eta_{ij}\psi_{ij} = \eta_{ij}\zeta_{ij}\psi_{ij} = \eta_{ij}[\G(\hat{\psi}_\rho)]_{ij}
\end{equation}
Therefore, also for the output state we can write $\G(\ro) = \hat\eta \circ \G(\hat{\psi}_\rho)$. The map $\mathcal{S}_\eta$ is a positive, trace preserving, unital map; we can again use the Uhlmann's majorization theorem and the Schur–concavity of passive state's energy to establish that the fixed-energy optimal input state for coherent ergotropy is pure. This motivates the choice of optimizing on pure states to evaluate the curves for the maximal coherent ergotropy in Fig.~\ref{fig:erg-beta}.

\section{Work Functionals Definitions}
\label{app:workfunc}

\subsection{Local Ergotropy}
The \textit{local ergotropy} quantifies the maximum amount of work which is \textit{locally unitarily extractable} from a system $\mathcal{H}_S$ in a state $\ro_S$, i. e. work extractable only using unitary transformations $\mathbb{U}_{loc}(d)$ that act locally on the subsystems of $\mathcal{H}_S$:
\begin{equation}
    \label{eq:def-ergloc}
    \erg(\ro; \hat{H}) \coloneqq \max_{\hat{U}\in \mathbb{U}_{loc}(d)}\left\{\tr[\ro \hat{H}]-\tr[\hat{U}\ro \hc{\hat{U}} \hat{H}]\right\}
\end{equation}
\subsection{Maximal Work Functionals}
Given a quantum channel $\Lambda$, fixing the input energy $E \geq 0$ and the number $n$ of copies of the system $\mathcal{H}_S$, one can define the following Maximal Work Functionals (MWF) \cite{QWC}:
\begin{itemize}
    \item\textbf{\textit{Shell-Maximal Work Functionals for noisy channels}}.
    The \textit{n-cells shell-maximal work functional} for a noisy channel is:
    \begin{equation}
            \overline{\W}^{(n)}(\Lambda;E) = \max_{\ro^{(n)}\,\in\,\mathfrak{S}^{(n)}_E}\ \W(\Lambda^{\otimes n}(\ro^{(n)});\hat{H}^{(n)})\ ;\\
    \end{equation}
    where $\W$ is a placeholder for the selected work functional and the energy shell $\mathfrak{S}^{(n)}_E$ is defined as
    \begin{equation}
        \mathfrak{S}^{(n)}_E :=  \left.\left\{ \ro^{(n)} \ \right|\ \tr[{\ro^{(n)}}\hat{H}^{(n)}]= E \right\}\ .
    \end{equation}
    \item\textbf{\textit{Ball-Maximal Work Functionals for noisy channels}}.
    The \textit{n-cells ball-maximal work functional} for a noisy channel is:
    \begin{equation}
        \begin{gathered}
            \W^{(n)}(\Lambda;E) = \max_{\ro^{(n)}\,\in\,\mathfrak{B}^{(n)}_E}\ \W(\Lambda^{\otimes n}(\ro^{(n)});\hat{H}^{(n)})\ ;
        \end{gathered}
        \label{eq:def_maxfunc}
    \end{equation}
    where again $\W$ stands for the selected work functional and the energy ball $\mathfrak{B}^{(n)}_E$ is defined as
    \begin{equation}
        \mathfrak{B}^{(n)}_E :=  \left.\left\{ \ro^{(n)} \ \right|\ \tr[{\ro^{(n)}}\hat{H}^{(n)}]\leq E \right\}\ .
    \end{equation}
\end{itemize}
It is evident from the definitions that
\begin{equation}
    \W^{(n)}(\Lambda;E) = \max_{0 \leq E' \leq E}\ \overline{\W}^{(n)}(\Lambda;E')\ .
    \label{eq:erg-is-max-of-barerg}
\end{equation}
In the operational scenario where only separable input states are admissible, one may define the following:

\textbf{\textit{Separable-Input Maximal Work Functionals for noisy channels}}. The \textit{n-cells separable-input maximal work functional} for a noisy channel is:
\begin{equation}
    {\W}^{(n)}_{sep}(\Lambda;E) = \max_{\ro^{(n)}\,\in\,\mathrm{SEP}[\mathfrak{I}^{(n)}_E]}\ \W(\Lambda^{\otimes n}(\ro^{(n)});\hat{H}^{(n)})\ ;\\
\end{equation}
where $\W$ is a placeholder for the selected work functional, the set $\mathfrak{I}^{(n)}_E$ is either $\mathfrak{S}^{(n)}_E$ or $\mathfrak{B}^{(n)}_E$ as defined before, and the optimization is performed over the sole separable states in the set, $\mathrm{SEP}[\mathfrak{I}^{(n)}_E]$.

\subsection{MAWERs}
A way of taking advantage of the $n$ copies of the system, which is alternative to what done to define work capacitances, is to use them to store a finite amount of energy $E$, such that $E/n = \e \to 0$ as $n\to\infty$. In this framework, it is appropriate to study the \textit{Maximal Asymptotic Work-Energy Ratio} (MAWER), defined as \cite{QWC,WEroleofcoherence}:
\begin{equation}
    {\J}_{\W}(\Lambda) := \limsup_{E \to \infty}\left(\sup_{n\geq\ceil*{E/\e_{max}}}\frac{\W^{(n)}(\Lambda;E)}{E}\right)\ ,
    \label{eq:def_MAWER1}
\end{equation}
where $\W$ is a placeholder for the selected work functional, and $\e_{max}$ is the greatest eigenvalue of the cell Hamiltonian. This definition is equivalent to \cite{WEroleofcoherence}:
\begin{equation}
    {\J}_{\W}(\Lambda) := \lim_{E \to \infty}\left(\lim_{n\to\infty}\frac{\W^{(n)}(\Lambda;E)}{E}\right)\ .
    \label{eq:def_MAWER2}
\end{equation}
Analogously, the separable-input MAWER is
\begin{equation}
    {\J}_{\W}(\Lambda) := \lim_{E \to \infty}\left(\lim_{n\to\infty}\frac{\W^{(n)}_{sep}(\Lambda;E)}{E}\right)\ .
    \label{eq:def_sep-MAWER1}
\end{equation}
Note that the order of the limits matters. Even if MAWERs and work capacitances refer to well defined and different use scenarios, a remarkable result from \cite{WEroleofcoherence} highlights a link between these quantities: given $\Lambda$ a LCPT map,
\begin{equation}
    \J_{\W}(\Lambda) = \lim_{\e\to 0}\frac{C_{\W}(\Lambda;\e)}{\e}\ .
    \label{eq:MAWER-derivative-WC}
\end{equation}
Thus, knowing $C_{\W}(\Lambda;\e)$ allows to easily derive the value of the associated MAWER.

\section{GMAD Channel Acting on a Qutrit}
\label{app:gmadqutrit}
Here we introduce the GMAD used in Secs.~\ref{sec:degco} and \ref{sec:enact}. Following Sec.~\ref{sec:defpro} we only need to choose the value of the coefficients of the unitaries $u_{ij}^{(2)}$ and $u_{ij}^{(1)}$. Having in mind a physical setting in which the system and environment interact via an energy preserving coupling for a given amount of time $t$, we will obtain the coefficients of interest by studying the evolution generated by the Hamiltonian $\hat{H}_{tot} = \hat{H}_S + \hat{H}_B + \hat{H}_I$ where 
\begin{gather}
    \hat{H}_{S} = \sum_{l=0}^{d-1} \epsilon_l\proj{l}_{S}\ , \qquad  \hat{H}_{B} = \sum_{k}\omega_{k}\proj{\omega_k}_{B}+\omega_{00}\proj{0}_B\ ,\\ 
    \hat{H}_{k} = g_k(\hat{\sigma}_{k,S}^{-}\ketbra{\omega_k}{0}_B + \hat{\sigma}_{k,S}^{+}\ketbra{0}{\omega_k}_B )\ ,\quad  \hat{H}_I = \sum_{k} \hat{H}_{k}\ .
    \label{eq:H-int-gmad}
\end{gather}
Here, the index $k$ runs over all the possible ordered pairs $\{r s\},\ r > s$, of levels of the qutrit, and the system jump operators are $\hat{\sigma}_{rs}^{-} \coloneqq \ketbra{s}{r}$, $\hat{\sigma}_{rs}^{+} \coloneqq \ketbra{r}{s}$.
The calculations of the Kraus operators $\hat{K}_{\omega_k\,,\, \omega_h} = \sqrt{\tau_h}\brab{\omega_k} \hat{U} \ketb{\omega_h}$ can be easily done after noting three fundamental properties
\begin{enumerate}
    \item Since the interaction Hamiltonian commutes with the free evolution, it is sufficient to evaluate $\brab{\omega_k}e^{-\im t \hat{H}_I}\ketb{\omega_h}$ instead of the full unitary generated by $\hat{H}_{tot}$.
    \item  the terms $\hat{H}_{ij}$ and $\hat{H}_{rs}$ commute for every $\{ij\}$ and $\{rs\}$, unless $s = j$:
\begin{equation}
    \begin{aligned}
        [\hat{H}_{ij},\hat{H}_{rs}] = g_{ij}g_{rs} \delta_{js}\left(\ketbra{i}{r}\otimes\ketbra{\omega_{ij}}{\omega_{rs}}-\ketbra{r}{i}\otimes\ketbra{\omega_{rs}}{\omega_{ij}}\right)\ ,
    \end{aligned}
    \label{eq:Hij-Hrs-do-not-commute}
\end{equation}
This relation is illustrated in Fig.~\ref{fig:commutation-gmad}.
\begin{figure}
    \centering
    \includegraphics[width=0.7\linewidth]{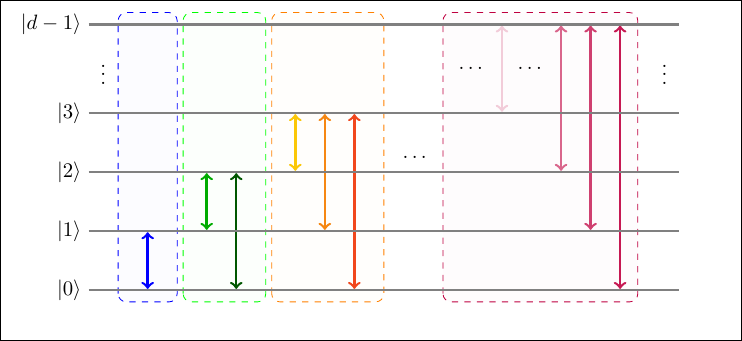}
    \caption{.The unitary evolution under $\hat{H}_I$ can be decomposed as a product of commuting terms (colored boxes); two terms $\hat{H}_{ij}$ and $\hat{H}_{rs}$ belonging to different boxes commute, while pairs from the same box do not commute.}
    \label{fig:commutation-gmad}
\end{figure}
Using equation \eqref{eq:Hij-Hrs-do-not-commute} it is possible to rewrite:
\begin{equation}
    \label{eq:UiProductDec}
    \begin{aligned}
        e^{-\im t \hat{H}_I} &= e^{-\im t \hat{H}_{01}}e^{-\im t (\hat{H}_{02}+\hat{H}_{12})}e^{-\im t (\hat{H}_{03}+\hat{H}_{13}+\hat{H}_{23})}...
    \end{aligned}
\end{equation}
    \item  The action of $\hat{H}_I$ can be decomposed on the different subspaces $\mathcal{H}_{SB}^{(E)}$, as discussed in Appendix \ref{app:microGMAD}. In particular, $\hat{H}_{I}$ acts trivially on the subspace spanned by the kets $\ket{0,\omega_0}$, $\ket{0,\omega_{21}}$, $\ket{1,\omega_{10}}$, $\ket{1,\omega_{20}}$, $\ket{2,\omega_{10}}$, $\ket{2,\omega_{21}}$, $\ket{2,\omega_{20}}$, while it acts as a Pauli ${\hat{\sigma}}_x$ on the subspace $\{\ket{0,\omega_{10}},\ket{1,\omega_{00}}\}$. In the three-dimensional subspace $\{\ket{0,\omega_{20}},\ket{1,\omega_{21}},\ket{2,\omega_{00}}\}$, $\hat{H}_{I}$ acts as a linear combination of the Gell-Mann matrices $\lambda_4$ and $\lambda_6\,$; indicating with $\tilde{H}$ the interaction Hamiltonian restricted on this subspace, we have
\begin{equation}
    \tilde{H} = 
    \begin{pmatrix}
        0 & 0 & g_{20}\\
        0 & 0 & g_{21}\\
        g_{20} & g_{21} & 0
    \end{pmatrix} 
    = g_{20}\lambda_4 + g_{21}\lambda_6\ .
\end{equation}
\end{enumerate}
Thus, we conveniently identify as natural candidates for the unitary matrices $u^{(1)}$ and $u^{(2)}$ the exponential of the matrices $(-it\,g_{10}\hat{\sigma}_x)$ and $(-it\tilde{H})$, which we can write explicitly using the standard Pauli exponentiation formula \cite{nielsen2010quantum} and the results in \cite{curtright2015elementary} for the exponentiation of Gell-Mann matrices. Hence, we have
\begin{align}
    u^{(1)} &= \cos (g_{10}t)\,\identity_2 - i \sin (g_{10}t)\, \hat{\sigma}_x =
    \begin{pmatrix}
        \cos (g_{10}t) & -\sin (g_{10}t) \\
        -\sin (g_{10}t) & \cos (g_{10}t)
    \end{pmatrix}\ ,
    \\
    u^{(2)} &= 
    \identity_3 - i\sin{(\bar{g}t)}\frac{\Tilde{H}}{\bar{g}} + (\cos{(\bar{g}t)}-1)\frac{\Tilde{H}^2}{\bar{g}^2} =
    \begin{pmatrix}
        1 + \frac{g_{20}^2}{\bar g^2}(\cos{(\bar{g}t)}-1) & \frac{g_{20}g_{21}}{\bar g^2}(\cos{(\bar{g}t)}-1) & -i \frac{g_{20}}{\bar g}\sin (\bar{g}t)\\
        \frac{g_{20}g_{21}}{\bar g^2}(\cos{(\bar{g}t)}-1) & 1 + \frac{g_{21}^2}{\bar g^2}(\cos{(\bar{g}t)}-1)& -i \frac{g_{21}}{\bar g}\sin (\bar{g}t)\\
        -i \frac{g_{20}}{\bar g}\sin (\bar{g}t) & -i \frac{g_{21}}{\bar g}\sin (\bar{g}t) & \cos{(\bar{g}t)}
    \end{pmatrix}\ ,
\end{align}
where $\bar{g} \coloneqq \sqrt{g_{20}^2 + g_{21}^2}$.
Therefore, it is possible to calculate the Kraus operators, which are listed below:
\begin{equation}
    \begin{aligned}
        &\hat{K}_{00}^{(D)} := r_{00}\,\mathrm{diag}[1,c_1,\bar{c}]\ ,
        \ \ 
        \hat{K}_{10}^{(D)} := r_{10}\,\mathrm{diag}[c_1,1,1]\ ,
        \\
        \\
        &\hat{K}_{21}^{(D)} := r_{21}\,\mathrm{diag}[1,\alpha_0^2+\alpha_1^2\bar{c},1]\ ,\ \ \hat{K}_{20}^{(D)} := r_{20}\,\mathrm{diag}[\alpha_1^2+\alpha_0^2\bar{c},1,1]\ ;
        \\
        \\
        &\hat{K}_{10}^{(1)} := -\im r_{00}s_1\ketbra{0}{1}\ ,
        \ \ 
        \hat{K}_{21}^{(2)} :=  -\im r_{00}\bar{s}\alpha_1\ketbra{1}{2}\ ,
        \\
        \\
        &\hat{K}_{20}^{(2)} :=  -\im r_{00}\bar{s}{\alpha_0}\ketbra{0}{2}\ ,
        \ \ 
        \hat{K}_{10}^{(2)} := r_{20}(\bar{c}-1){\alpha_0\alpha_1}\ketbra{0}{1}\ ;
        \\
        \\
        &\hat{K}_{01}^{(1)} := \frac{r_{10}}{r_{00}}{\hat{K}_{10}^{(1)\,\top}}\ ,\ \ \hat{K}_{12}^{(2)} := \frac{r_{21}}{r_{00}}{\hat{K}_{21}^{(2)\,\top}}\ ,\ \ \hat{K}_{02}^{(2)} := \frac{r_{20}}{r_{00}}{\hat{K}_{20}^{(2)\,\top}}\ ,\ \ \hat{K}_{01}^{(2)} := \frac{r_{20}}{r_{21}}{\hat{K}_{10}^{(2)\,\top}}\ ;
        \end{aligned}
    \label{eq:gmad-2-qutrit-K-operators}
\end{equation}
where the coefficients $r_{ij}$ are defined in Eq. \eqref{eq:unitdb} and $c_1,s_1,\bar{c},\bar{s},\alpha_0,\alpha_1$ are real numbers derived from the coupling constants $g_{k}$ in Eq. \eqref{eq:H-int-gmad} as follows:
\begin{equation}
    \begin{aligned}
        &c_1 \coloneqq \cos{(g_{10}t)}\ ,\ \ s_1 \coloneqq \sin{(g_{10}t)}\ ,
        \\
        &\bar{c} \coloneqq \cos{(\bar{g}t)}\ ,\ \ \bar{s} \coloneqq \sin{(\bar{g}t)}\ ,
        \\
        &\alpha_0 \coloneqq g_{20}/\bar{g}\ ,\ \ \alpha_1 \coloneqq g_{21}/\bar{g}\ .
    \end{aligned}
    \label{eq:plotparameters}
\end{equation}
Here the time interval $t$ is fixed, each value of $t$ corresponding to a different set of values for the entries of the unitary matrices; it could also be normalized away by conveniently choosing an appropriate unit of measure for the time (or equivalently for the couplings $g$).
\\
In the limit of vanishing temperature, discarding all the terms proportional to $e^{-\beta\omega}$ for any $\omega>0$, one retrieves the MAD channel output matrix of \cite{chessa2021quantum}, upon identifying:
\begin{equation}
    \g_1' := \sin^2(g_{10}t)\ ;\quad
    \g_2' := \frac{g_{21}^2}{\bar{g}^2}\sin^2(\bar{g}t)\ ;\quad
    \g_3' := \frac{g_{20}^2}{\bar{g}^2}\sin^2(\bar{g}t)\ .
    \label{eq:B-to-MAD-mapping}
\end{equation}
with $\g$ used according to the notation of \cite{chessa2021quantum}. Thus, defining a \textit{$\beta$-family} the set of channels $\G$ with fixed couplings $g_{ij}$ but various $\beta$, $\{\G_{\beta}\}_{\beta}$, one can associate to every $\beta$-family of channels $\G$ a MAD channel $\mathcal{M}$ through the mapping in \eqref{eq:B-to-MAD-mapping}. By virtue of this and the other many properties shared by the GMAD and MAD channels, it is reasonable to propose $\G$ as a faithful generalization of $\mathcal{M}$ at non-zero temperature.

\section{Coherent EME in GMADs}
\label{app:cEMEtab}
Here we show that GMAD channels are potentially capable of originating a coherent ergotropic Mpemba effect by explicitly providing a channel and a pair of states to observe the effect:
\begin{equation}
    \begin{aligned}
        \ro_1 &= \begin{pmatrix}
        0.61419885 & 0.23993793 -0.20486506i & 0.14709909 +0.05359668i\\
        0.23993793 +0.20486506i & 0.22372136 & 0.04471051 +0.09323805i\\
        0.14709909 -0.05359668i & 0.04471051 -0.09323805i & 0.16207979
    \end{pmatrix}\ ,
    \\
    \ro_2 &= \begin{pmatrix}
        0.61419885 & 0.02280222 -0.30977726i & -0.08634611 +0.05914107i\\
        0.02280222 +0.30977726i & 0.22372136 &
               0.04784115 +0.00286678i\\
        -0.08634611 -0.05914107i & 0.04784115 -0.00286678i & 0.16207979
    \end{pmatrix}\ ,
    \end{aligned}
\end{equation}
with a GMAD such that $g_{10} = 0.8$, $g_{21} = 0.2$, $g_{20} = 0.1$, $\beta = 0.1$, and system Hamiltonian with spectrum $(0,0.5,1)$.
\end{widetext}
\bibliography{biblio_file}

\end{document}